\documentclass{iopart}
\usepackage{epsfig}
\usepackage{bm}

\def\Journal#1#2#3#4#5{#1 {\it #2} #3 {\bf #4} #5}
\def\Preprint#1#2{#1 {\it Preprint} #2}

\def\EPJ{Eur. Phys. J.}
\def\HIP{Heavy Ion Phys.}
\def\NP{Nucl. Phys.}
\def\PL{Phys. Lett.}
\def\PR{Phys. Rev.}
\def\PRL{Phys. Rev. Lett.}

\newcommand{\xo}{x_{\rm out}}
\newcommand{\xs}{x_{\rm side}}

\newcommand{\lapp}{\,{\raisebox{-.2ex}{$\stackrel{<}{_\sim}$}}\,}
\newcommand{\half}{{\textstyle{\frac{1}{2}}}}

\begin{document}

\title{Hydrodynamic emission of strange and non-strange
particles at RHIC and LHC}

\author{Ulrich Heinz\dag\footnote[3]{email: heinz@mps.ohio-state.edu} 
}
\address{\dag\ 
Physics Department, The Ohio State University, Columbus, OH 43210}

\begin{abstract}
The hydrodynamic model is used to describe the single-particle spectra
and elliptic flow of hadrons at RHIC and to predict the emission angle
dependence of HBT correlations at RHIC and LHC energies.
\end{abstract}

\section{Introduction and Summary}

One of the key signatures of the formation of thermalized hot and
dense matter, such as a quark-gluon plasma (QGP), in relativistic
heavy-ion collisions is collective transverse flow, driven by the
transverse pressure gradients between the center of the hot collision
fireball and the surrounding vacuum. The space-time evolution of this 
collective flow can be described by relativistic hydrodynamics. In the 
limit of zero mean free path (i.e. when microscopic scattering rates 
exceed any macroscopically relevant time scale) the system can be 
described as an ideal fluid neglecting transport effects such as 
viscosity and heat conduction. The flow observed in the final state
via analysis of the single-particle spectra of emitted hadrons can 
then be directly related to the initial conditions of the collision
fireball when it first thermalized, and to the equation of state $p(e,n_B)$ 
of the expanding matter. 

In non-central collisions the initial conditions are anisotropic, due to 
the spatial deformation of the nuclear overlap region in the transverse 
plane, resulting in anisotropic pressure gradients which ultimately lead 
to anisotropies in the final momentum distributions. The momentum 
anisotropies can be quantified by the coefficients $v_n$ of an azimuthal 
Fourier decomposition of the transverse momentum distributions. These 
coefficients, in particular the {\em elliptic flow coefficient} $v_2$,
grow only as long as anisotropies in the spatial pressure gradients
persist. Since the reaction zone quickly evolves from its initial
spatially deformed state towards a more cylindrical shape without
significant anisotropies, the growth of $v_2$ is restricted to the 
earliest stages of the fireball expansion \cite{Sorge97}. Even in 
the absence of interactions among the fireball constituents the 
spatial anisotropy which could drive the build-up of elliptic flow 
decays quickly, due to radial free-streaming of the matter \cite{KSH00}. 
Delayed thermalization therefore manifests itself in reduced elliptic 
flow since $v_2$ then has to build on smaller anisotropies in the 
pressure gradients. If thermalization happens quickly, elliptic flow 
builds up earlier and reaches larger values. The resulting stronger
flow into the reaction plane \cite{Ollitrault92} destroys the spatial
deformation even faster than free-streaming, resulting in an early
saturation of the momentum anisotropy after about 5 fm/$c$ in 
semiperipheral Au+Au collisions at RHIC \cite{KSH00}. For a given 
initial deformation (i.e. a fixed impact parameter) the saturation
value of $v_2$ has been shown to be a monotonic function of the 
microscopic scattering cross section \cite{ZGK99}, reaching a strict
upper limit given by ideal fluid hydrodynamics in the limit of zero
mean free path. The fact that the elliptic flow measured at RHIC
almost exhausts this hydrodynamic limit (for an extensive list of 
references I refer to the recent review \cite{Kolb:2003dz}) is at this
point the strongest available evidence for early thermalization (at 
$\tau_{\rm therm}<1$\,fm/$c$ \cite{HK02}) and, together with an
estimated energy density $>10$\,GeV/fm$^3$ at this early time
\cite{KSH00}, strongly suggests the formation of a well-thermalized 
QGP with significant (5--7\,fm/$c$) lifetime \cite{HK02}.

We have demonstrated elsewhere \cite{Kolb:2003dz,HK02} that the bulk
of the observed hadrons from Au+Au collisions at RHIC (i.e. all mesons
in the region $p_{\rm T} \lapp 1.5-2$\,GeV/$c$ and all baryons at  
$p_{\rm T} \lapp 2-2.5$\,GeV/$c$) behave hydronymically, i.e. that 
their single particle spectra and momentum anisotropies in central
and semicentral collisions with impact parameter $b\lapp 7$\,fm can all
be quantitatively reproduced by relativistic ideal fluid hydrodynamics
with a common kinetic decoupling temperature. In Section 2 I supplement 
this finding by pointing out that the same holds true even for the heavy
$\Omega$ and $\bar\Omega$ hyperons, in contrast to earlier expectations
based on their presumed weaker coupling to the expanding pion fluid during 
the late hadronic expansion stage \cite{HSN98}. It has also been noted 
\cite{HK02,HK02WWND} that the hydrodynamic model does much more poorly
in its description of the two-particle Bose-Einstein (Hanbury Brown Twiss
or HBT) correlations, by overpredicting the longitudinal and outward
HBT radii while underpredicting the sideward radius and its transverse
momentum dependence (the ``RHIC HBT puzzle''). In Section 3 I show how
HBT studies as function of the azimuthal emission angle can constrain
the space-time evolution of the {\em spatial} deformation in 
non-central collisions, thereby supplementing the information from the
measured momentum-space anisotropies and constraining the total duration 
of the expansion phase. Preliminary RHIC data from 2-pion correlations
\cite{Lopez02,Lisa03} agree with the hydrodynamic prediction 
\cite{HK02HBTosci} that at kinetic freeze-out the fireball is (on 
average) still slightly out-of-plane elongated, although much less so 
than initially. This is expected to change at LHC energies where the
much higher initial energy density allows the elliptic flow to act for 
a significantly longer time period until decoupling, leading to pion 
emission from an (on average) in-plane elongated source. The HBT signatures
of such an in-plane elongation are discussed in Section 3.

\section{Radial flow of $\Omega$ hyperons}

\begin{figure}[ht]
\centerline{\epsfig{file=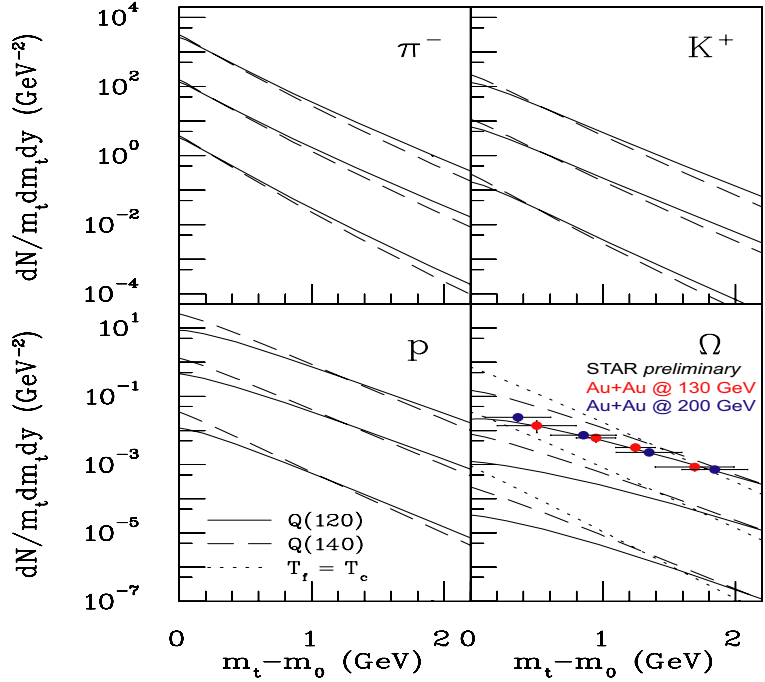,height=9cm,width=10cm,angle=0}}
\caption{Hydrodynamically predicted hadron spectra from 
Ref.~\protect\cite{HKHRV01}, together with preliminary $\Omega{+}\bar\Omega$
spectra from the STAR collaboration~\protect\cite{STAR03omega}.
See text for details.
\label{F1}
}
\end{figure}

When shortly after the beginning of the experimental program at RHIC in
2000 the first data on the total charged particle multiplicity in central
Au+Au became available \cite{PHOBOS02dN130}, Peter Kolb, Pasi Huovinen 
and I used this to constrain the initial conditions for the hydrodynamic
model in central conditions and to predict hadron spectra and their 
anisotropies for non-central collisions \cite{HKHRV01}. Some of these 
predictions are shown in Fig.~\ref{F1}. The three sets of curves 
correspond, from top to bottom, to central ($b{\,<\,}5.4$\,fm), 
semiperipheral ($5.4\,{\rm fm}{\,<\,}b{\,<\,}9.9$\,fm) and peripheral 
($9.9\,{\rm fm}{\,<\,}b{\,<\,}13.5$\,fm) Au+Au collisions at 
$\sqrt{s}=130\,A$\,GeV. At that time no spectra were available yet to
fix the freeze-out temperature, so we estimated it to probably lie
between 120 and 140 MeV (solid and dashed curves, respectively). 
We later learned \cite{HK02} that $T_{\rm f}=130$\,MeV fits the pion 
and proton data from central collisions best. For the $\Omega$ hyperons
in the lower right panel we also included a prediction for freeze-out
directly after hadronization at $T_{\rm c}=165$\,MeV, motivated by the
prevailing prejudice \cite{HSN98} that, due to their lack of resonant 
scattering with pions, they would not be able to efficiently pick up
%
%
any additional transverse flow generated in the hadronic phase. This
expectation was supported by observations by the WA97 Collaboration at 
the SPS \cite{Antinori:2000sb} which indicated a smaller slope parameter
for $\Omega$ and $\bar\Omega$ than for other hadrons. It was only recently
realized that this evidence was at least partially distorted by fitting 
a curved $\Omega$ spectrum in a limited $m_{\rm T}$ region
by a simple exponential. Recent NA49 data for $\Omega$ and $\bar\Omega$
at smaller $m_{\rm T}$ show evidence that the spectra bend over at
low $m_{\rm T}$ and seem to be consistent with much larger radial flow
than previously thought, almost as strong as that experienced by pions and
nucleons \cite{Afanasiev:2002fk}. 

Figure~\ref{F1} shows that the same is true at RHIC: The $\Omega$ spectra 
from Au+Au collisions at $\sqrt{s}{\,=\,}130$ and 200\,$A$\,GeV 
\cite{STAR03omega} are consistent with the hydrodynamic predictions if 
they are assumed to freeze out at the same decoupling temperature 
$T_{\rm f}\simeq130$\,MeV as pions, kaons and (anti)protons, while 
they are much too flat to be compatible with freeze-out already at 
$T_{\rm c}$, i.e. directly after hadronization. It is impossible to 
generate enough radial flow for the $\Omega$s already before 
hadronization, without destroying the agreement of the model with
all other hadron spectra and/or lowering the hadronization temperature
significantly below 165\,MeV (which would contradict lattice QCD results).
We conclude that the $\Omega$ hyperons seem to be much more strongly 
coupled to the expanding pion fluid during the late hadronic stage
than previously thought. It will be interesting to explore the microscopic 
mechanisms responsible for such a strong coupling.

\section{Emission angle dependent HBT interferometry}

The observational constraints on dynamical models for the evolution 
of the heavy-ion reaction zone can be further tightened by a 
simultaneous analysis of the single-particle momentum spectra with 
2-particle momentum correlations. The latter provide access to the 
space-time structure of the source at kinetic freeze-out. More 
specifically, in the case of Bose-Einstein correlations between 
identical bosons, the Gaussian widths $R^2_{i,j}(\bm{K})$ of the 
correlation function $C(\bm{q},\bm{K}) = 1 + \lambda(\bm{K})
\exp\bigl[-\sum_{i,j=o,s,l}q_iq_jR^2_{i,j}(\bm{K})\bigr]$ are linear
combinations of the components of the spatial correlation 
tensor $S_{\mu\nu}(\bm{K}) = \bigl[\langle x_\mu x_\nu\rangle-
\langle x_\mu\rangle\langle x_\nu\rangle\bigr](\bm{K})$ which 
characterize the space-time widths of the source emission function
$S(x,K)$ at freeze-out. For the general case these relations can be found 
in \cite{Wiedemann98,HHLW02}; for longitudinally boost-invariant sources
and pions at midrapidity $Y=0$ they simplify to 
\begin{eqnarray}
\label{equ:HBTbneq0}
R_s^2(\Phi)    &=&   \half (S_{xx}+S_{yy}) 
                   - \half (S_{xx}-S_{yy}) \cos 2 \Phi      
                   - S_{xy} \sin 2 \Phi                              \\
R_o^2(\Phi)    &=&   \half (S_{xx}+S_{yy}) 
                   + \half (S_{xx}-S_{yy}) \cos 2 \Phi 
                   + S_{xy} \sin 2 \Phi                 \nonumber    \\
               & & -2 \beta_{\rm T} (S_{tx} \cos \Phi + S_{ty} \sin \Phi)
                   + \beta_{\rm T}^2 S_{tt}                                \\
R_{os}^2(\Phi) &=&   S_{xy} \cos 2 \Phi
                   - \half (S_{xx}-S_{yy}) \sin 2 \Phi  \nonumber    \\
               & & + \beta_{\rm T}(S_{tx} \sin 2\Phi - S_{ty} \cos \Phi)   \\
R_l^2   (\Phi) &=&   S_{zz}\,. \label{equ:HBTbneq0last} 
\end{eqnarray}
For simplicity the dependence of the spatial correlation tensor $S_{\mu\nu}$
on the pair momentum $\bm{K}$ (in particular its {\em implicit} dependence
on the azimuthal emission angle $\Phi$ of $\bm{K}$ relative to the
reaction plane) has been dropped, and only the {\em explicit} 
$\Phi$-dependence arising from the azimuthal rotation between the
outward direction $\xo \parallel \bm{K}_{\rm T}$ and the reaction plane
coordinate $x\parallel\bm{b}$ is exhibited. For small transverse pair
momenta $K_{\rm T}$ this explicit $\Phi$-dependence dominates \cite{LHW00}.
The source widths $S_{\mu\nu}$ are specified in coordinates $(x,y)$ 
parallel and perpendicular to the reaction plane while the correlation 
function is measured in $(\xo,\xs)$ coordinates defined by the directions
parallel and perpendicular to $\bm{K}_{\rm T}$. $\beta_{\rm T}$ is the 
transverse pair velocity. The terms $R_{ol}^2$ and $R_{sl}^2$ vanish
due to the assumed longitudinal boost-invariance.   

The orientation of the reaction plane in the laboratory system can be 
determined event-by-event from the elliptic flow. The correlation 
function can thus be measured as a function of the azimuthal emission 
angle $\Phi$ relative to the reaction plane. For non-central collisions 
the spatial deformation of the source in the transverse plane leads to
oscillations as a function of $\Phi$. In practice the finite angular bin
size and reaction plane resolution smear these azimuthal oscillations and
reduce their measured amplitudes, but this can be fully corrected for 
in a model-independent way \cite{HHLW02}. General symmetries
ensure that the HBT radii extracted from a Gaussian fit to the thus 
corrected correlation function exhibit the following general 
$\Phi$-dependences \cite{HHLW02}:
\begin{eqnarray}
 \label{24}
   R_s^2(\Phi) &=& R_{s,0}^2 + {\textstyle2\sum_{n=2,4,6,\dots}} 
   R_{s,n}^2\cos(n\Phi),
 \nonumber\\
   R_o^2(\Phi) &=& R_{o,0}^2 + {\textstyle2\sum_{n=2,4,6,\dots}} 
   R_{o,n}^2\cos(n\Phi),
 \nonumber\\
   R_{os}^2(\Phi) &=& \qquad\;\;
   {\textstyle2\sum_{n=2,4,6,\dots}} R_{os,n}^2\sin(n\Phi),
 \nonumber\\
   R_l^2(\Phi) &=& R_{l,0}^2 + {\textstyle2\sum_{n=2,4,6,\dots}}
   R_{l,n}^2\cos(n\Phi).
 \end{eqnarray}
For sources without longitudinal boost-invariance $R^2_{ol}$ and
$R^2_{ol}$ oscillate around zero with cosines and sines, respectively,
of {\em odd} multiples of $\Phi$ \cite{HHLW02,LHW00}. The relations
(\ref{24}) continue to hold true for correlation measurements made 
in a finite symmetric window around $Y{\,=\,}0$. 

%
\begin{figure}[ht]
\centerline{\epsfig{file=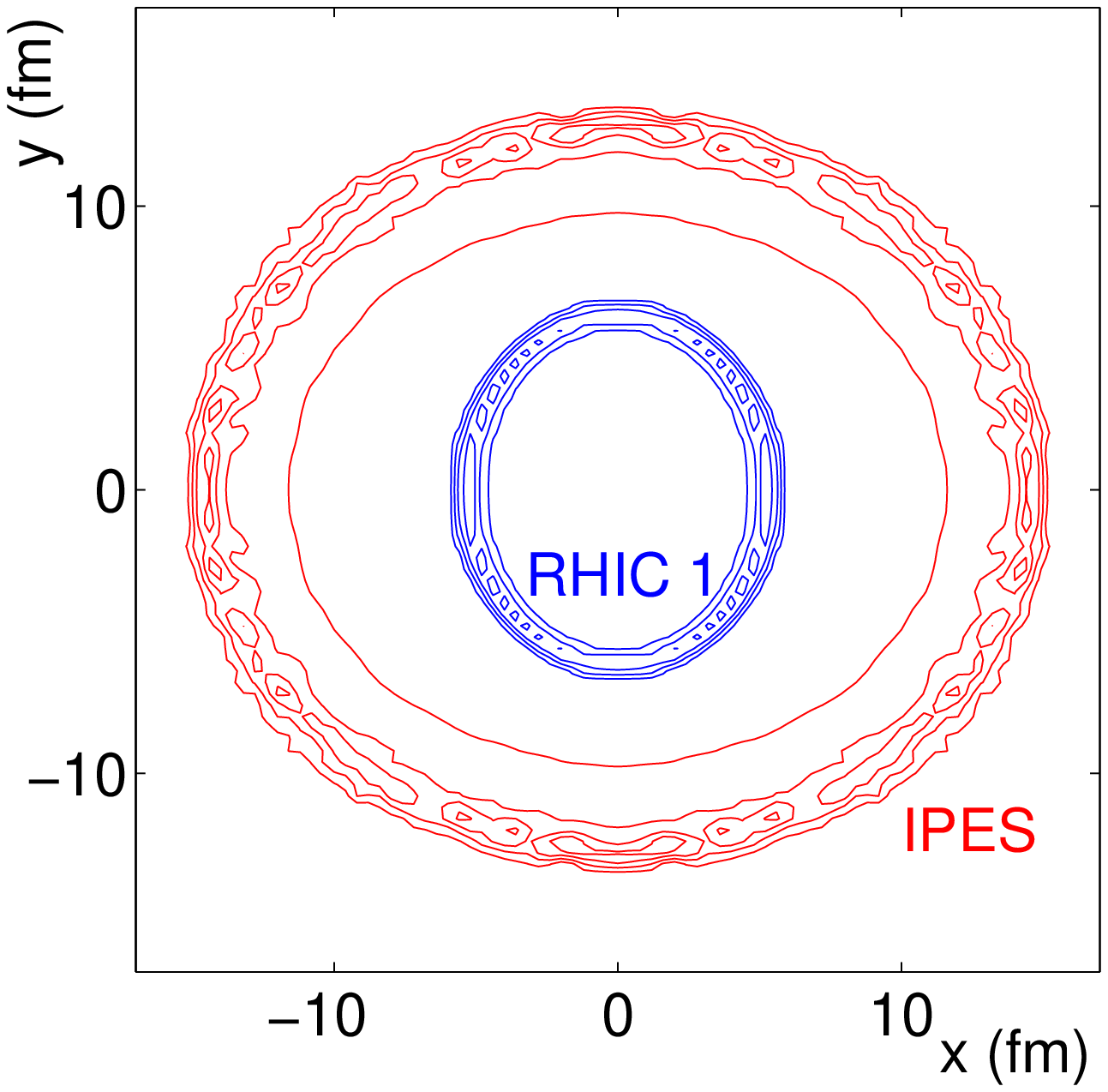,height=5cm,angle=0}
            \epsfig{file=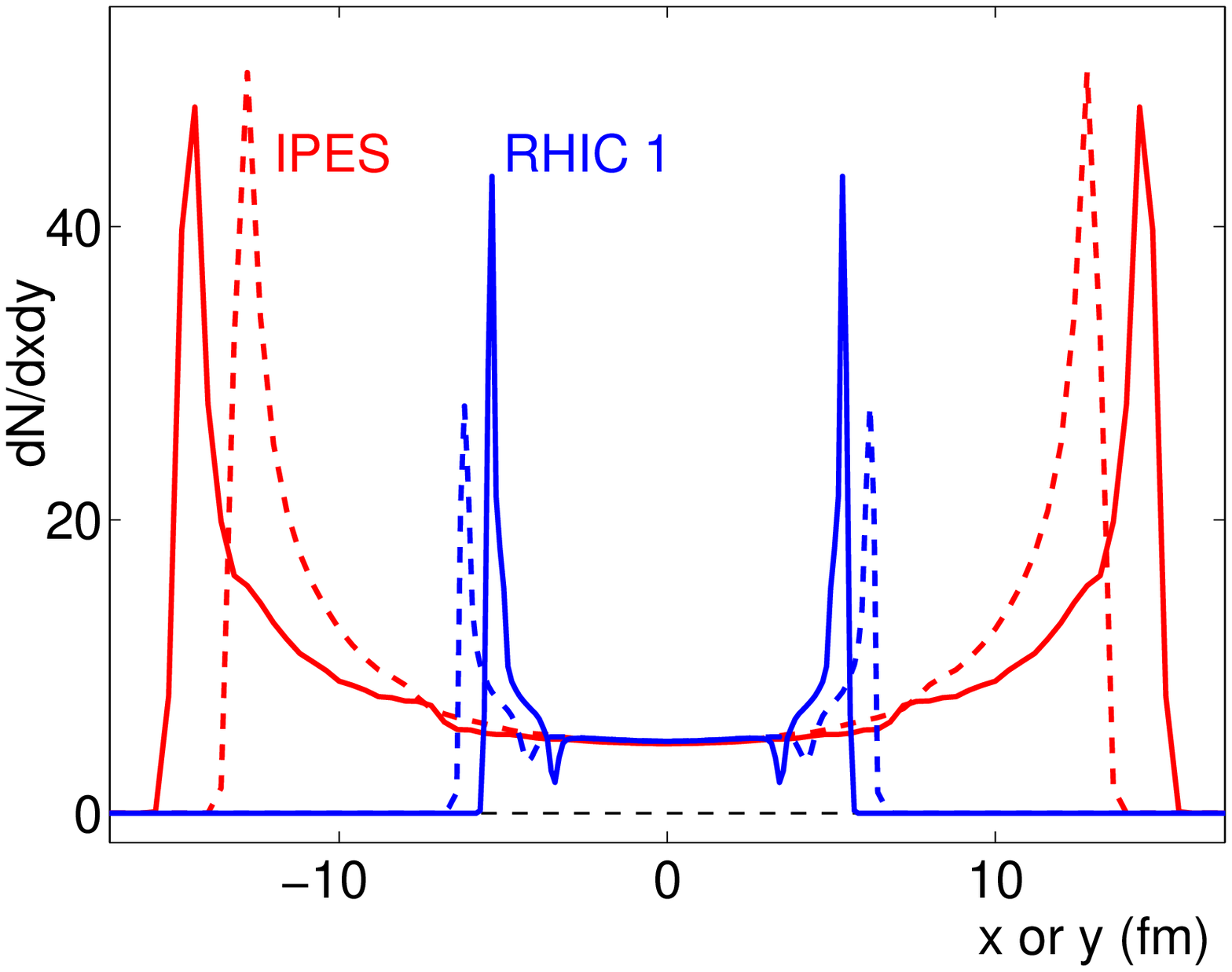,height=5.01cm,width=6cm,angle=0}}
\caption{Contour plots of and horizontal (solid) and vertical (dashed)
cuts through the $\bm{K}$-integrated pion emission function
in the transverse plane for RHIC- and LHC-typical initial
conditions, for Au+Au collisions at $b{\,=\,}7$\,fm 
\protect\cite{HK02HBTosci}.
\label{F2}
}
\end{figure}
%
Figure~\ref{F2} shows the momentum-integrated hydrodynamic emission 
function for RHIC- and LHC-type initial conditions. The curves labelled 
``RHIC\,1'' correspond to the same initial conditions which led to a
successful description of the single-particle momentum spectra and elliptic 
flow in 130\,$A$\,GeV Au+Au collisions. One sees that, when integrated
over all momenta and all emission times, this source is still out-of-plane 
elongated, although only very slightly. ``IPES'' labels an in-plane-elongated
source obtained by starting the hydrodynamics earlier at a higher
initial temperature of 2\,GeV \cite{HK02HBTosci}. In this case the
source lives longer, grows much bigger and changes its initial 
out-of-plane elongation into a significant in-plane elongation
at freeze-out. 

The vertical and horizontal cuts in the right panel of Fig.~\ref{F2} show 
that both sources are very ``opaque'' in that most particles are emitted 
from the (deformed) surface. A differential analysis shows 
\cite{HK02HBTosci} that only pions with small transverse momenta come 
from the interior of the source; they are thus the only ones which probe 
the overall spatial deformation of the source in the transverse plane. 
Pions with non-zero transverse momenta, on the other hand, come from
thin slivers near the edge of the fireball in the emission direction
(for details see \cite{HK02HBTosci}); they do not directly probe the
overall shape of the source.

%
\begin{figure} 
\centerline{
            \epsfig{file=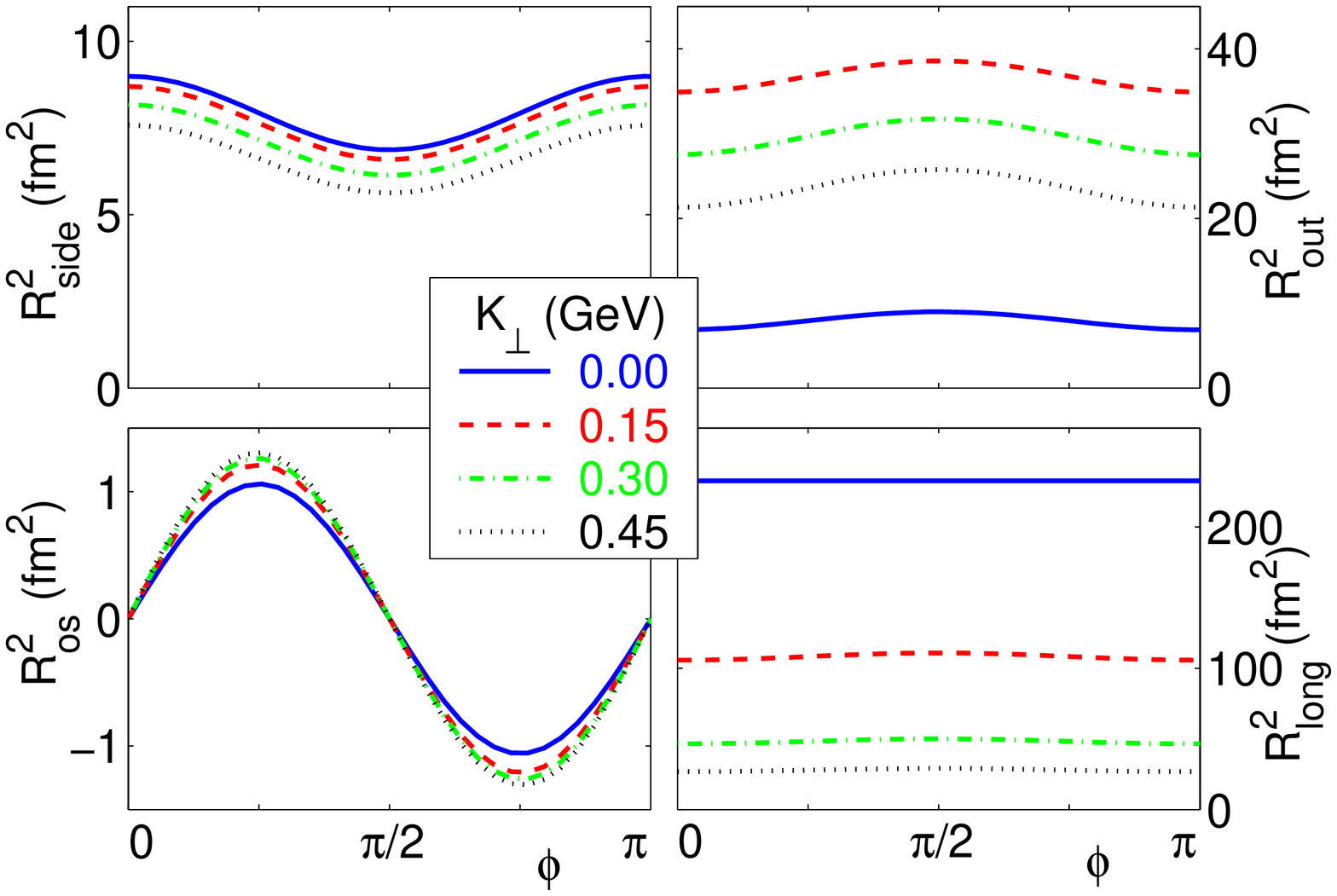,width=6cm} \hfill
            \epsfig{file=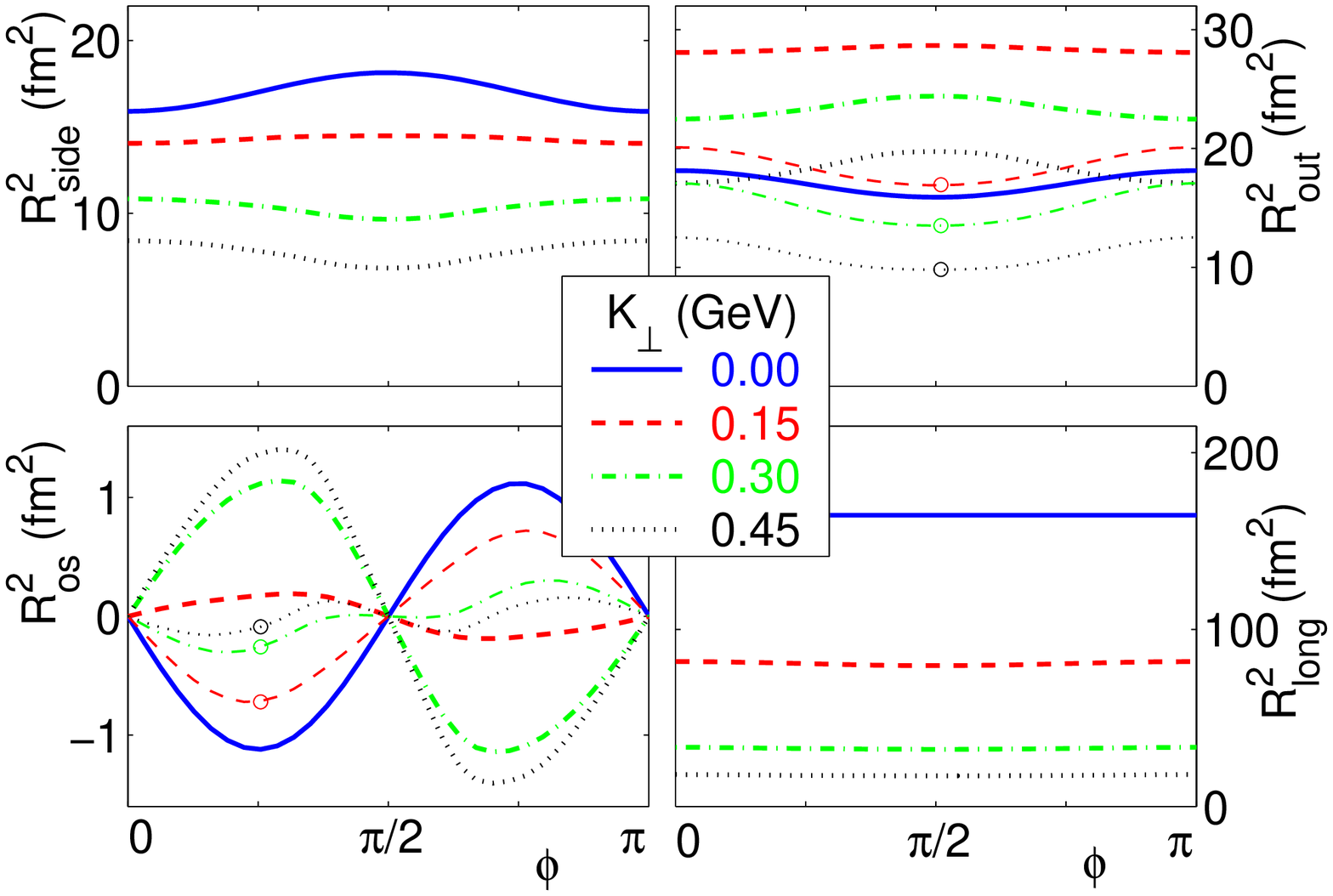,width=6cm}
            }  
\caption{Oscillations of the HBT radii for different transverse 
         pair-momenta in RHIC collisions (left) and for the source 
         elongated into the reaction plane (right) \protect\cite{HK02HBTosci}.
         The thin circled lines in the right panel show the geometric 
         contributions to the HBT radius parameters. 
\label{F3} 
} 
\vspace*{-3mm}
\end{figure} 
%
Figure~\ref{F3} shows the azimuthal oscillations of the four 
non-vanishing HBT radius parameters for several values of $K_{\rm T}$, 
calculated from these emission functions.
The opposite spatial deformation of the RHIC\,1 and IPES sources is
reflected in the opposite signs of the oscillations for $R_s^2$, 
$R_o^2$ and $R_{os}^2$ at $K_{\rm T}{\,=\,}0$. 
For example, at RHIC energies $R_s^2$ oscillates downward, implying a 
larger sideward radius when viewed from the $x$ direction (i.e. within 
the reaction plane) than from the $y$-direction (i.e. perpendicular to 
the reaction plane).
For the in-plane elongated source (IPES) some of the oscillation amplitudes
change sign at larger transverse momenta.
This change of sign originates from an intricate interplay between geometric,
dynamical and temporal aspects of the source at freeze-out \cite{HK02HBTosci}
and reinforces the statement that at $K_{\rm T}\ne0$ the HBT correlations
do {\em not} probe the entire source and its spatial shape.
Preliminary STAR data on azimuthal oscillations of the HBT radii
\cite{Lopez02,Lisa03} confirm the oscillation pattern in the left panel 
of Fig.~\ref{F3} but disagree with the one shown in the right panel.
This suggests that hydrodynamics gets the overall duration of the
expansion stage about right, since significantly longer times until 
freeze-out should cause the source to reverse the sign of its spatial
deformation. 
More quantitative statements have to await a resolution of the 
above-mentioned ``HBT puzzle'' in central collisions.
The significance of the $K_{\rm T}$-dependence of the oscillation 
amplitudes for the HBT radius parameters is still largely unexplored. 
From the analysis of central collisions \cite{TWH99} it is known that
the $\Phi$-averaged transverse HBT radii (in particular $R_o$) exhibit 
stronger $K_{\rm T}$ dependence for sources with a sharp surface (such 
as the hydrodynamic one obtained from the Cooper-Frye prescription) than
for Gaussian sources.
One might expect similarly strong differences for the $K_{\rm T}$-dependence
of the oscillation amplitudes in non-central collisions.\\[-2mm]

\noindent{\bf Acknowledgement:} This work was supported by the 
U.S. Department of Energy under Grant No. DE-FG02-01ER41190.
\vspace*{-2mm}
\section*{References}
{}
\end{document}